\def\<{\langle}
\def\>{\rangle}

\documentstyle[aps,prl,multicol,epsf,epsfig,float]{revtex}

\title{An Analysis of a Two-Atom Double-Slit Experiment
Based on \\ Environment-Induced Measurements}
\author{Christian Sch\"on$^{1,2}$ and Almut Beige$^{1,2}$}
\address{$^1$Optics Section, Blackett Laboratory, Imperial College
London, London SW7 2BZ, England. \\
$^2$Max-Planck-Institut f\"ur Quantenoptik, Hans-Kopfermann-Str. 1,  85748 
Garching, Germany.\footnote{Present address.}}
\date{\today}

\begin{document}

\maketitle
\draft

\begin{abstract}
\begin{center}
\parbox{14cm}
{To investigate the effect of the environment on a quantum 
mechanical system we consider two two-level atoms 
in a free radiation field in the presence of a screen. By assuming that 
the screen causes {\em continuous ideal measurements} on the free radiation field 
we derive a quantum jump description for the state of the atoms. 
Our results are consistent with the master equation for {\em dipole}
interacting atoms, but give more insight in the time evolution of a
{\em single} system. To illustrate this we derive a necessary and sufficient 
criterion for interference in a two-atom double-slit experiment
and analyse bunching in the statistics of photons emitted in a certain 
direction.}
\end{center} 
\end{abstract} 

 \vspace*{0.2cm}
\noindent

\begin{multicols}{2}

\section{Introduction}

In this paper we study the effect of the environment on a simple quantum 
mechanical system. The experimental setup we consider as an example 
is shown schematically in Fig.~\ref{exp}. It consists of two two-level atoms 
continuously driven by a resonant laser field  and stored at a fixed 
distance $r$ from each other. The atoms are surrounded by a free radiation field
and spontaneously emit photons. Each photon causes a ``click'' 
at a certain point on a screen. If enough photons are emitted, 
these ``clicks'' add up and form an interference pattern. 

\noindent
\begin{minipage}{3.38truein}
\begin{center}
\begin{figure}[h]
\epsfig{file=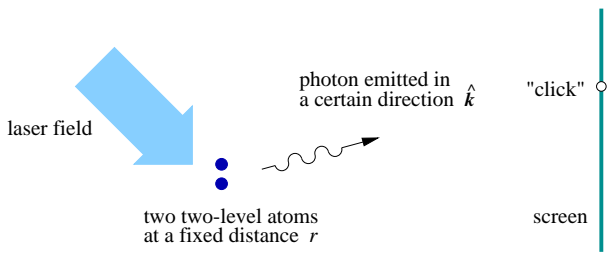,width=8cm}\\[0.1cm]
\caption{
Experimental setup. Two two-level atoms are placed at a fixed distance $r$
from each other and are continuously driven by a resonant laser. This leads to 
spontaneous photon emissions. Each photon causes a ``click" on a screen
in a direction $\hat{\bf k}$ away from the atoms.}
\label{exp}
\end{figure}
\end{center}
\end{minipage}

\vspace*{0.3cm}

The Hamiltonian $H$ of the quantum mechanical system, which consists here of the 
two atoms, the laser and the free radiation field, is well known \cite{meystre}.
However, solving the corresponding Schr\"odinger equation does not explain that
the atoms emit spontaneously photons. On the other hand, a purely wave 
mechanical description of the emitted photons can predict the interference 
pattern \cite{Heitler,Mandel} 
but does not allow us to determine higher-order time correlations 
in the photon statistics.

The aim of this paper is to show that the experiment pictured in Fig.~\ref{exp} 
can be explained purely quantum mechanically from first principles 
with the help of the projection 
postulate for ideal measurements \cite{PP}. We show that the 
environment surrounding the system -- the screen -- has the same effect 
as continuous measurements on the free radiation field.
That each photon causes a ``click'' on the screen at a point that depends 
only on the direction of its wave vector ${\bf k}$ suggests that the screen measures 
whether a photon has been emitted or not. If so it determines its direction 
$\hat{\bf k} = {\bf k}/k$. As these measurements are caused
by the interaction of the free radiation field with the screen, we call them 
{\em environment induced measurements}.

Between consecutive measurements the state of the atoms and the field develops 
with the Hamiltonian $H$ and all components of the quantum mechanical system 
become entangled. A measurement on the free radiation field therefore also has
an effect on the atomic state. In case of a ``click'' on the screen 
the state of the atoms changes abruptly. 
It {\em jumps} into the reset state which can be obtained by applying
the {\em reset operator} $R_{\hat{\bf k}}$ to the state $|\psi\rangle$ 
of the two atoms before the emission. 

By deriving the reset operator $R_{\hat{\bf k}}$ \cite{Molmer} we specify the quantum jump approach 
for two {\em dipole} interacting atoms \cite{Beige,Beige2} which predicts the 
no photon time evolution with the help of the conditional Hamiltonian $H_{\rm cond}$ 
but does not distinguish between photon emissions in different directions $\hat{\bf k}$.
To justify the assumptions and approximations on which our results are based
we show that they are consistent with the master equation for two dipole 
interacting atoms \cite{Dicke,Lehmberg,Agarwal}. Both approaches, the quantum jump 
approach and the master equation, are widely used in quantum optics and both have their 
respective merits. 

A quantum jump description 
\cite{Hegerfeldt92,Dalibard,Carmichael,Hegerfeldt93} is well suited 
for predicting all possible trajectories of a {\em single} system. 
Using this approach, it has been shown, 
for instance, that {\em environment induced measurements} can assist 
in the realisation of universal gates for quantum computing \cite{PRL}. 
A possible application of the reset operator 
$R_{\hat{\bf k}}$ is given by a recently proposed scheme by Cabrillo {\em et al.}
\cite{Cabrillo} for entangling distant atoms by interference.
The master equation has considerable advantages in the description of 
an {\em ensemble} of systems and are well suited for determining stationary states.

The main reason to consider in this paper an experimental setup with {\em two} 
atoms is that this leads to spatially dependent effects which 
do not occur in single atom experiments. Verifying these effects experimentally
shows that the quantum jump approach is not only an artifact of the master 
equations obtained from an unraveling of these equations \cite{ravel} but 
a self-consistent approach. The aim of this paper is to show that the quantum jump 
approach can be applied to all experiments 
in which a single system spontaneously emits photons and is surrounded by
``white'' walls of a laboratory forming the screen.

The experimental setup shown in Fig.~\ref{exp} has been discussed widely 
in the literature \cite{Mandel,Fano,Richter91,Huang,Rudolph,Brewer}
and it has been realised as a quantum mechanical two-atom double-slit experiment
by Eichmann {\em et al.}~\cite{Eichmann} in 1993.
The slits of the classical version of this experiment are there replaced 
by two atoms which are likewise the sources of the light reaching the screen. 
In spite of its simplicity and the fact that this experiment is one of the 
basic experiments in quantum mechanics its discussion never came to an end. 
For other recent and related quantum mechanical double-slit experiments see 
Refs.~\cite{Grangier,Kwiat92,Duerr,Kwiat99,Kim}.

Here we show, in agreement with Refs.~\cite{Wong,Skornia}, that the reset operator 
$R_{\hat{\bf k}}$ allows us to determine {\em directly} the interference pattern of
the experiment by Eichmann {\em et al.}~\cite{Eichmann}. To demonstrate the advantage 
of the quantum jump approach we derive a {\em necessary and sufficient} criterion for 
interference. In good agreement with Refs.~\cite{Kwiat92,Itano,Dirac,Feynman},
it is shown that 
interference arises from the fact that in quantum mechanics the wave functions, and not 
the probabilities, of different paths contributions have to be added to determine the 
probability for a certain event to happen. 
Other authors attributed interference in quantum mechanical double-slit experiments
to the position-momentum uncertainty relation, Bohr's complementarity principle 
and to the absence of the {\em which way} information 
\cite{Duerr,Scully82,Scully91,Storey,Englert95,Wiseman,Englert96,Knight,Jakob,Mahler,Dung}.
It is shown here for the experimental setup of Fig.~\ref{exp}
that the interference vanishes if and only if the which way 
information is, at least in principle, available in the experiment. 

To give a further application of our results we analyse the effect of 
bunching in the statistics of photons emitted in a certain direction 
$\hat{\bf k}$. In agreement with Ref.~\cite{Wiegand,Skornia} 
we predict arbitrary strong bunching even if the atoms are several wave-lengths 
apart from each other. An intuitive explanation for this effect is 
given following the reasoning of Ref.~\cite{Beige}. 

This paper is organised as follows. In Section II we derive the reset operator 
$R_{\hat{\bf k}}$ which represents the main result of our paper.
In Section III we give a short overview of the quantum jump approach
and show its consistency with the master equation 
for two dipole interacting atoms \cite{Agarwal}. In Section IV we 
discuss the experimental setup shown in Fig. 1 and 
derive a necessary and sufficient interference criterion. Afterwards we discuss 
spatially dependent bunching in the statistics of the photons emitted by the two atoms.
Finally, our results are summarised in Section VI.

\section {The reset operator}

In this section we derive an analytic expression for the {\em reset operator} 
$R_{\hat{\bf k}}$ which can be used to determine the state of the atoms after an 
emission in a certain direction $\hat{\bf k}$ from first principles. 
If the state of the atoms just before an emission is $|\psi\rangle$ it can, 
as we show below, immediately afterwards be written as
\begin{eqnarray} \label{23}
|\hat{\psi}_{\hat{\bf k}}\> & \equiv & 
R_{\hat{\bf k}} |\psi\>/\|\cdot \| ~,
\end{eqnarray}
which is a pure state. This equation defines the operator $R_{\hat {\bf k}}$ up to a 
proportionality factor. For practicality we choose 
this factor such that the probability density for a photon emission in the
$\hat{\bf k}$ direction, $I_{\hat{\bf k}}(\psi)$, equals
\begin{eqnarray} \label{I-R}
I_{\hat{\bf k}}(\psi) & \equiv & \| R_{\hat{\bf k}} |\psi\> \|^2 ~,
\end{eqnarray}
which is a density in time and solid angle.

To derive an analytic expression for the reset operator let us first write down 
the Hamiltonian of the quantum mechanical system consisting of two two-level atoms 
and the free radiation field. In the following 
$|1\>_i$ and $|2\>_i$ denote the ground state and the excited state of atom $i$ and
$S^-_i=|1\>_{ii}\<2|$ and $S^+_i=|2\>_{ii}\<1|$ are the corresponding 
lowering and raising operators. The energy separation between the levels is given by 
$\hbar \omega_0$. The annihilation operator for a single photon of the 
mode $({\bf k}, \lambda)$ of the free radiation field is denoted by 
$a_{{\bf k}\lambda}$ where ${\bf k}$ is its wave vector, $\lambda$ characterises 
its polarisation and ${\bf \epsilon}_{{\bf k}\lambda}$ is the polarisation vector. 
The coupling constant between the free radiation field and atom $i$ is given by 
$g^{(i)}_{{\bf k}\lambda}$. For simplicity we assume that both atoms have the 
same transition dipole moment ${\bf D}_{21}$ which gives 
$g^{(1)}_{{\bf k}\lambda}=g^{(2)}_{{\bf k}\lambda} = g_{{\bf k}\lambda}$ with
\begin{eqnarray}
g_{{\bf k}\lambda} 
&=& {\rm i} e \, \left( \frac{\omega_k}{2 \epsilon_0 \hbar L^3 } \right)^{1/2}
\! {\bf D}_{21} \cdot {\bf \epsilon}_{{\bf k}\lambda} ~,
\end{eqnarray}
where $\omega_k=k/c$ and $L^3$ is the quantisation volume. 
In addition, we assume that both atoms are irradiated by a laser field which
has the (complex) Rabi frequency $\Omega^{(i)}$ with respect to atom $i$. 
If both atoms interact with the same laser the relative phase of the two 
Rabi frequencies depends on the direction of the incoming beam.
Using this notation the interaction Hamiltonian $H_{\rm I}$ with 
respect to the interaction-free Hamiltonian is given by
\begin{eqnarray} \label{21}
H_{\rm I} &=& \hbar \sum_{i=1,2} \sum_{{\bf k}, \lambda} 
              {\rm e}^{{\rm i} (\omega_0-\omega_k) t} \,
              {\rm e}^{{\rm i} {\bf k} \cdot {\bf r}_i} \, 
              g_{{\bf k}\lambda} \, a_{{\bf k}\lambda} S^+_i  + {\rm h.c.} \nonumber \\
& & + {\hbar \over 2} \sum_{i=1,2} \Omega^{(i)} \,S^+_i + {\rm h.c.}
\end{eqnarray}

In the experimental setup of Fig.~\ref{exp}, 
each emitted photon causes a ``click'' at a certain {\em point} on the screen. 
To describe this we assume that the presence of the screen leads to
repeated measurements on the free radiation field as to whether a photon has been 
emitted or not. If so it determines its direction $\hat{\bf k}$.
Here we do not discuss what exactly causes these
{\em environment induced measurements} but show later that the results 
derived from this assumption are consistent with the master equation for two 
dipole interacting atoms \cite{Agarwal} and in good agreement with 
the experimental results of Ref.~\cite{Eichmann}. To determine the state 
of the atoms in the case of a ``click'' 
we make use of the projection postulate for ideal measurements \cite{PP}. 

Let us first consider a situation in which the screen is 
replaced by detectors which measure with each photon also its wave vector ${\bf k}$
and polarisation $\lambda$. As in Refs.~\cite{Hegerfeldt93}, we assume
that the atoms are initially in state $|\psi\rangle$ and the free radiation 
field is in the vacuum state $|0_{\rm ph} \>$.  After a time $\Delta t$, which 
should not be too long so that in $\Delta t$ only the one-photon states become populated,
the detector performs a measurement on the free radiation field.
According to the projection postulate the unnormalised state 
of the atom-field system in the case of a ``click'' caused by a 
photon $|1_{{\bf k}\lambda} \rangle$ equals
\begin{eqnarray} \label{25}
|1_{{\bf k} \lambda}\> |\psi_{{\bf k} \lambda} \>
& \equiv & |1_{{\bf k}\lambda} \rangle \langle 1_{{\bf k}\lambda}|
U_{\rm I} (\Delta t,0) |0_{\rm ph} \> |\psi \> ~.
\end{eqnarray}
Here $U_{\rm I}(\Delta t,0)$ is the time development operator with 
respect to the interaction Hamiltonian (\ref{21}) which entangles
the state of the atoms with the state of the free radiation field. 
The measurement of the free radiation field therefore also has an effect on the 
atomic state. It makes the atoms {\em jump} into the state 
$|\psi_{{\bf k} \lambda} \>$.

A comparison of both sides of Eq.~(\ref{25}) shows that 
the unnormalised state of the atoms after the ``click'' of the detector equals
\begin{eqnarray} \label{26}
|\psi_{{\bf k} \lambda} \>
& = & \< 1_{{\bf k} \lambda}| U_{\rm I} (\Delta t,0) |0_{\rm ph} \> |\psi\> ~.
\end{eqnarray}
From first order perturbation theory and Eq.~(\ref{21}) we find 
\begin{eqnarray} \label{27}
|\psi_{{\bf k} \lambda} \>
& = & - {\rm i} \, g_{{\bf k} \lambda}^* \, 
\int_0^{\Delta t} {\rm d}t \, {\rm e}^{-{\rm i} (\omega_0-\omega_k)t} 
\sum_{i=1,2} {\rm e}^{ -{\rm i}{\bf k} \cdot {\bf r}_i }\,S^-_i |\psi \> ~. 
\nonumber \\ &&
\end{eqnarray}
According to the projection postulate \cite{PP}, the squared norm of this vector
equals the probability density for the emission of a photon $|1_{{\bf k}\lambda} \rangle$ 
during the time interval $\Delta t$. Assuming $\Delta t \gg 1/\omega_0$ we obtain 
in analogy to Refs.~\cite{Hegerfeldt93}
\begin{eqnarray} \label{huhuuu}
I_{{\bf k}\lambda}(\psi) 
&=& \lim_{\Delta t \to 0}
{\| \, |\psi_{{\bf k} \lambda}\> \|^2 \over \Delta t} \nonumber\\ 
&=& 2\pi \, |g_{{\bf k}\lambda}|^2 \, \delta(\omega_0 - \omega_k) \, 
\Big\| \sum_{i=1,2} {\rm e}^{ -{\rm i}{\bf k} \cdot {\bf r}_i }\,S^-_i |\psi\> \Big\|^2 ~.
\nonumber \\ &&
\end{eqnarray}
The proportionality of this equation to $\delta(\omega_0 - \omega_k)$ shows that
all emitted photons have, within the approximations made, the wave number $k_0=\omega_0 c$. 
The normalised state of the atoms after an emission therefore equals
\begin{eqnarray} \label{29}
|\hat{\psi}_{\hat{\bf k}} \>
& = & \Big( \sum_{i=1,2} {\rm e}^{ -{\rm i} k_0 \, {\hat{\bf k}} \cdot {\bf r}_i }
\, S^-_i |\psi \> \Big)/\| \cdot \| ~,
\end{eqnarray}
which depends only on the direction $\hat{\bf k}$ of the emitted photon 
but not on $k$ and $\lambda$.

Let us now consider again the situation where each emitted photon is detected 
by a ``click'' on the screen which determines only its direction $\hat{\bf k}$. 
To find the state of the atoms after an emission in this case we can proceed as 
above but have to replace the projector 
$|1_{{\bf k}\lambda} \rangle \langle 1_{{\bf k}\lambda}|$ in Eq.~(\ref{25}) by  
\begin{eqnarray} 
I\!\!P_{\hat{\bf k}} 
& = & \sum_{k,\lambda} |1_{k \, \hat{\bf k} \lambda} \rangle 
\langle 1_{k \, \hat {\bf k} \lambda}|~.
\end{eqnarray}
This operator projects onto all one-photon states with a wave vector in the $\hat{\bf k}$
direction. By doing so we find that the reset state of the atom-field system
equals 
\begin{eqnarray} \label{299}
\sum_{k,\lambda} |1_{k \, \hat{\bf k} \lambda}\> |\psi_{k \, \hat{\bf k} \lambda} \> 
& \equiv & \sum_{k,\lambda} |1_{k \, \hat{\bf k}\lambda} \rangle 
\langle 1_{k \, \hat{\bf k}\lambda}| U_{\rm I} (\Delta t,0) |0_{\rm ph} \> |\psi \> ~.
\nonumber \\ &&
\end{eqnarray}
As shown above, only terms with $k=k_0$ contribute with a non-vanishing 
amplitude to the right hand side of this equation. From Eq.~(\ref{29})
one can then see that Eq. (\ref{299}) is of the form
\begin{eqnarray} 
\sum_{k,\lambda} |1_{k \, \hat{\bf k} \lambda}\> |\psi_{k \, \hat{\bf k} \lambda} \> 
&=& \sum_{\lambda} c_\lambda \, |1_{k_0 \, \hat{\bf k} \lambda}\> 
|\hat{\psi}_{\hat{\bf k}} \> ~,
\end{eqnarray}
where $c_\lambda$ is a complex number.
Normalising this state we find that $|\hat{\psi}_{\hat{\bf k}} \>$ of Eq.~(\ref{29})
is indeed the reset state (\ref{23}) of the atoms.

The probability density for a ``click'' on the screen in the direction $\hat{\bf k}$ 
away from the atoms can be obtained from the relation
\begin{eqnarray} \label{211}
I_{\hat{\bf k}}(\psi) 
& = & \sum_{\lambda} \left( {L \over 2 \pi} \right)^3 \int_0^\infty {\rm d}k \, k^2 \, 
I_{k \, \hat{\bf k} \lambda}(\psi) ~.
\end{eqnarray}
Using Eq.~(\ref{huhuuu}) this leads to
\begin{eqnarray} \label{212}
I_{\hat{\bf k}}(\psi) 
& = & \frac{3 A}{8 \pi} \left(1- |{\bf D}_{21} \cdot \hat{\bf k}|^2 \right) 
\Big\| \sum_{i=1,2} {\rm e}^{ -{\rm i}k_0 \, {\hat{\bf k}} \cdot {\bf r}_i }\,S^-_i |\psi\> \Big\|^2 ~,
\nonumber \\ &&
\end{eqnarray}
where
\begin{eqnarray} \label{213}
A &=& \frac{e^2 \omega_0^3 \, |{\bf D}_{21}|^2}{3\pi \epsilon_0 \hbar c^3}
\end{eqnarray}
is the spontaneous decay rate of a single atom.

From Eq.~(\ref{29}) and (\ref{212}) we can now derive an expression
for the reset operator $R_{\hat{\bf k}}$ of Eq.~(\ref{23}) and (\ref{I-R}) and find
\begin{eqnarray} \label{reset}
R_{\hat{\bf k}} &=& R_{\hat{\bf k}}^{(1)} + R_{\hat{\bf k}}^{(2)} 
\end{eqnarray}
with
\begin{eqnarray} \label{215}
R_{\hat{\bf k}}^{(i)} &=& 
\left[ {3A \over 8\pi} \left( 1 - |{\hat{\bf D}}_{21} \cdot \hat{\bf k}) |^2 \right)
\right]^{1/2} {\rm e}^{-{\rm i} k_0 \, {\hat{\bf k}} \cdot {\bf r}_i} \, S^-_i ~.
\end{eqnarray}
In the same way as shown here for two atoms, one can derive the reset operator 
for the situation when only atom $i$ is emitting photons whilst the other atom is 
far away and cannot emit a photon onto the same point on the screen. 
Proceeding as above we find that the reset operator in this case 
is given by $R_{\hat{\bf k}}^{(i)}$ of Eq.~(\ref{215}) alone. The reset 
operator for {\em both} atoms is the sum of the reset operators for each individual 
atom. This fact will play an important role in the discussion of a two-atom 
double-slit experiment in Section IV.

\section{Quantum jump approach versus master equation}

Before we apply our results to the experimental setup of Fig.~\ref{exp} we 
shortly summarise the quantum jump approach \cite{Review}
and show that the results obtained in the Section II are consistent with the
master equation for two dipole interacting atoms \cite{Agarwal}. 

\subsection{The quantum jump approach}

The quantum jump approach 
\cite{Hegerfeldt92,Dalibard,Carmichael,Hegerfeldt93} can be used to 
predict all possible trajectories of a single quantum mechanical system 
which stochastically emits photons. At all times $t$ the probability density 
for a photon emission is known. If this happens the state of the atoms 
changes abruptly. It jumps into another state 
which can be determined with the help of the reset operator. Between two 
photon emissions the system undergoes a continuous time 
evolution which can be described by the conditional Hamiltonian
$H_{\rm cond}$. 

To derive $H_{\rm cond}$ for two dipole interacting atoms one can
proceed as in Section II. Assuming again that the environment performs 
repeated measurements on the free radiation field one can determine the state 
of the system in the case of no photon emission by replacing the projector 
$|1_{{\bf k}\lambda} \rangle \langle 1_{{\bf k}\lambda}|$ in Eq.~(\ref{25}) by  
the projector onto the vacuum state 
$|0_{\rm ph} \rangle \langle 0_{\rm ph}|$. In this way one finds that the state 
of the atom-field system equals in the case of no photon emission 
after a time interval $\Delta t$
\begin{eqnarray} \label{u} 
|0_{\rm ph} \>U_{\rm cond}(\Delta t,0) |\psi \> 
& \equiv & |0_{\rm ph} \rangle \langle 0_{\rm ph}| U_{\rm I} 
(\Delta t,0) |0_{\rm ph}\> |\psi \> ~.
\nonumber \\ &&
\end{eqnarray}
Using second order perturbation theory this leads, as in Ref.~\cite{Beige}, to
\begin{eqnarray} \label{hc}
H_{\rm cond} & = & \frac{\hbar}{2{\rm i}} 
\Big[ \, A \sum_{i=1,2} S_i^+ S_i^- 
+ C \sum_{i \neq j} S_i^+S_j^- \, \Big] \nonumber \\
& & + {\hbar \over 2} \sum_{i=1,2} \Omega^{(i)} \,S^+_i + {\rm h.c.}
\end{eqnarray}
with the complex dipole interaction coupling constant
\begin{eqnarray} \label{C}
C & = & {3A \over 2} \, {\rm e}^{{\rm i} k_0 r} \,
\left[ \, \frac{1}{{\rm i} k_0 r} \left(1- | \hat{\bf D}_{21} \cdot \hat{\bf r} |^2 \right) 
\right. \nonumber\\
&   &  \left.+\left(\frac{1}{(k_0r)^2}-\frac{1}{{\rm i} (k_0r)^3} \right) 
\left(1- 3 \, | \hat{\bf D}_{21} \cdot \hat{\bf r} |^2 \right) \, \right] ~. 
\end{eqnarray}
As in Section II, we assume here that the dipole moment ${\bf D}_{21}$ is the same
for both atoms. 

The probability for no photon emission in $\Delta t$ can be obtained from 
Eq.~(\ref{u}) by taking the norm squared and equals
\begin{equation} \label{P0}
P_0(\Delta t,\psi) = \|  U_{\rm cond}(\Delta t,0) \, |\psi\>  \|^2 ~.
\end{equation}

\subsection{Consistency with the master equation for two dipole interacting atoms}

Another way to describe two atoms inside a free radiation field is to use the 
master equation. It provides linear differential equations which govern the time 
evolution of the density matrix $\rho$ corresponding to an ensemble of 
single systems. It can be derived by averaging over all possible trajectories. 
By doing so we show here that our results are consistent the with master equation 
for two dipole interacting atoms.

Let us now consider an ensemble of systems with initial state $\rho$. 
After a time $\Delta t$ this ensemble consists of many subensembles. 
The subensemble without photon emissions develops with the conditional Hamiltonian 
$H_{\rm cond}$ and can, at time $\Delta t$, be described by the density matrix 
\begin{equation}
\rho_0(\Delta t) = U_{\rm cond}(\Delta t,0) \rho U_{\rm cond}^\dagger(\Delta t,0) ~.
\end{equation}
Eq.~(\ref{P0}) shows that the trace over this matrix is equal to the probability for 
no photon emission in $(0,\Delta t)$ and to the relative size of the subensemble
without photon emissions. Using Eq.~(\ref{23}) and (\ref{I-R}) we see 
that the density matrix of the subensemble of systems with a photon emission 
in $\hat{\bf k}$ direction equals 
\begin {equation}
\rho_{\hat{\bf k}} \Delta t = R_{\hat{\bf k}} \rho R_{\hat{\bf k}}^\dagger \Delta t 
\end {equation}
and the trace over this matrix gives the relative size of this subensemble.

If $\Delta t$ is not too long so that the probability for more than one emission 
can again be neglected, the density matrix of the whole ensemble
at $\Delta t$ equals
\begin{equation} \label{Delta}
\rho(\Delta t) = \rho_0(\Delta t) + \sum_{\hat{\bf k}} \rho_{\hat{\bf k}} \Delta t~.
\end{equation}
From Eq.~(\ref{215}), (\ref{hc}) and (\ref{C}) we find
\begin{eqnarray} \label{sum}
\sum_{\hat{\bf k}} \rho_{\hat{\bf k}} 
&=& (A+{\rm Re}\,C)\,R_+\rho R_+^{\dagger} +(A-{\rm Re}\,C)\,R_-\rho R_-^{\dagger}
\end{eqnarray}
with
\begin{eqnarray}
R_\pm &=& (S_1^- \pm S_2^-)/\sqrt{2} ~.
\end{eqnarray}
Considering $\Delta t$ as a continuous parameter this leads to the differential 
equation
\begin {eqnarray}\label{master}
\dot{\rho} &=& 
-\frac{\rm i}{\hbar} \left[ H_{\rm cond} \,\rho - \rho\, H_{\rm cond}^\dagger 
\right] \nonumber\\
& & +(A + {\rm Re} \, C) \, R_+ \rho R_+^\dagger 
+ (A - {\rm Re} \, C) \, R_- \rho R_-^\dagger ~.
\end{eqnarray}
A comparison with Ref.~\cite{Agarwal} 
shows that this is the master equation for two dipole interacting 
atoms. 

\section{Analysis of the two-atom double-slit experiment}

To demonstrate the usefulness of the reset operator $R_{\hat{\bf k}}$ 
we apply it in this section to the two-atom double-slit experiment shown 
in Fig.~\ref{exp}. A necessary and sufficient criterion for 
interference is derived. The interference pattern we predict has the same 
spatial dependence as the one observed experimentally by 
Eichmann {\em et al.} \cite{Eichmann}.

\subsection{A necessary and sufficient criterion for interference}

Before we discuss the two-atom double-slit experiment in which the atoms
are continuously driven by a laser field, let us first consider a simplified 
version of the setup shown in Fig.~\ref{exp}. We assume that 
the atoms are repeatedly prepared in the same pure state $|\psi \rangle$. 
By observing the emitted photons one can
measure the spatially dependent probability density 
$I_{\hat{\bf k}}(\psi)$.

To calculate $I_{\hat{\bf k}}(\psi)$ we determine first 
the unnormalised reset state $|\psi_{\hat{\bf k}} \rangle$ of the two atoms in 
the case of an emission in the $\hat{\bf k}$ direction. From Eq.~(\ref{23}) and 
(\ref{reset}) we find that it is a superposition of two wave functions, each 
corresponding to a different situation, but both leading to a ``click'' at the 
same point on the screen,
\begin{eqnarray} \label{reset-state}
|\psi_{\hat{\bf k}} \rangle &=&
R_{\hat{\bf k}}^{(1)}|\psi\rangle + R_{\hat{\bf k}}^{(2)}|\psi\rangle ~.
\end{eqnarray}
The amplitude $R_{\hat{\bf k}}^{(i)}|\psi\rangle$ 
describes the state of the atoms after a photon emission by atom $i$ alone.
We denote the probability density for such an emission by 
$I_{\hat{\bf k}}^{(i)}(\psi)$. Analogously to Eq.~(\ref{I-R}) it equals
\begin{eqnarray} \label{ii}
I^{(i)}_{\hat{\bf k}}(\psi) &=& \| R_{\hat{\bf k}}^{(i)} |\psi\rangle \|^2 ~.
\end{eqnarray}
The probability density $I_{\hat{\bf k}}(\psi)$ can be obtained by taking the 
squared norm of the reset state $|\psi_{\hat{\bf k}} \rangle$ and we find 
\begin{eqnarray}
I_{\hat{\bf k}}(\psi) \label{i}
&=& I^{(1)}_{\hat{\bf k}}(\psi) + I^{(2)}_{\hat{\bf k}}(\psi)
+ 2 \, {\rm Re} \, \langle\psi| R_{\hat{\bf k}}^{(2)\dagger} R_{\hat{\bf k}}^{(1)} 
|\psi\rangle ~. \nonumber \\ &&
\end{eqnarray}
This differs by the last term from the sum of the probability densities for an 
emission either by atom 1 or atom 2 and describes the {\em interference} in the 
light emitted by the two atoms quantitatively. Interference results from the 
joint coupling of both atoms to the same free radiation field.
There is only no interference iff the last term in Eq.~(\ref{i})
vanishes for all directions $\hat{\bf k}$, i.e. 
\begin{eqnarray} \label{crit}
{\rm Re} \, \langle\psi| R_{\hat{\bf k}}^{(2)\dagger} R_{\hat{\bf k}}^{(1)} 
|\psi\rangle &=& 0 ~~ {\rm for~all} ~~ \hat{\bf k} ~.
\end{eqnarray}
This condition is equivalent to the reset states $R_{\hat{\bf k}}^{(1)} |\psi\rangle$ 
and $R_{\hat{\bf k}}^{(2)} |\psi\rangle$ being orthogonal to each other
and we find using Eq. (\ref{215}) that
\begin{eqnarray} \label{crit2}
\langle \psi| S_2^+ S_1^- |\psi\rangle & \neq & 0 
\end{eqnarray}
is a necessary and sufficient criterion for interference. Whether this criterion 
is fulfilled or not depends only on the initial state $|\psi\rangle$ of the atoms.

Summarising this, we have shown that interference in the two-atom double-slit 
experiment can be attributed to the fact that the amplitudes of the wave function 
corresponding to a ``click'' at the same point on the screen have to be added to 
determine the probability for this to happen. 
This is opposed to classical probability theory where the probabilities of all 
contributing paths have to be added, and which would not yield the last term in Eq.~(\ref{i}). 
Attributing interference to the superposition of wave functions is one of the
basic concepts in quantum mechanics \cite{Dirac,Feynman}. 
However, the quantum jump approach allowed us to calculate the amplitudes of the wave 
function for the concrete experimental setup shown in Fig.~\ref{exp} explicitly 
and to identify each amplitude with a certain path.

\subsection{The which way information}

Other authors showed that interference in quantum mechanical double-slit 
experiments vanishes in the presence of the which way information  
(see for instance Scully and Dr\"uhl \cite{Scully82}).
Englert \cite{Englert96} derived an inequality which relates the fringe 
visibility to the which-way knowledge available in 
the experiment. In the following, we show that this is in good 
agreement with the criterion given in Eq.~(\ref{crit2}). 

To do so we first point out that a which way interpretation 
automatically implies the assumption that each photon is emitted either by 
atom 1 or by atom 2. Assuming this, the quantum jump approach predicts that
the reset state of the atoms for a certain emission equals 
$R_{\hat{\bf k}}^{(i)} |\psi\rangle$ 
with the corresponding probability density 
$\|R_{\hat{\bf k}}^{(i)} |\psi\rangle\|^2$ where $i$ equals $1$ or $2$. 
This is in contradiction with Eq.~(\ref{i}) which shows that the probability 
density for an emission in the $\hat{\bf k}$ direction equals 
$\|R_{\hat{\bf k}} |\psi\rangle\|^2$ and not 
$\|R_{\hat{\bf k}}^{(1)} |\psi\rangle\|^2 + \|R_{\hat{\bf k}}^{(2)} |\psi\rangle\|^2$.

Nevertheless, there is one situation in which one cannot 
distinguish whether both atoms are cooperatively emitting or whether 
one can assign each photon to one of the two atoms. This is the case 
iff 
\begin{eqnarray} \label{perp}
R_{\hat{\bf k}}^{(1)} |\psi\rangle
& \perp & R_{\hat{\bf k}}^{(2)} |\psi\rangle ~~ {\rm for~all} ~~ \hat{\bf k} ~.
\end{eqnarray}
Then one can find out which atom emitted the photon by measuring whether the atoms are 
either in the state $R_{\hat{\bf k}}^{(1)} |\psi\rangle$ or 
in $R_{\hat{\bf k}}^{(2)} |\psi\rangle$.
Eq.~(\ref{215}) shows that Eq.~(\ref{crit}) and (\ref{perp}) are equivalent. 
This means, the interference vanishes if and only if the which way 
information is available in the experiment.

\subsection{Interference from two continuously driven atoms}

In the previous two subsections we assumed that the state of the atoms by the time 
of an emission is always $|\psi \rangle$. This is not the case for 
the experimental setup of Fig.~\ref{exp} in which the atoms are continuously 
driven by a laser field. To apply our results to this situation we have to 
describe the atoms at the time of an emission by the steady state matrix 
$\rho^{\rm ss}$. From Eq.~(\ref{I-R}) we find that the probability density for 
an emission in the $\hat{\bf k}$ direction equals
\begin{eqnarray} \label{intensity}
I_{\hat{\bf k}}(\rho^{\rm ss}) 
&=& {\rm Tr} \big( R_{\hat{\bf k}} \rho^{\rm ss} R_{\hat{\bf k}}^{\dagger} \big) ~.
\end{eqnarray}
In analogy to Eq.~(\ref{crit2}) a necessary and sufficient criterion for interference 
is now given by the condition
\begin{eqnarray} \label{crit3}
{\rm Tr} \big( S_2^+ S_1^- \rho^{\rm ss} \big) 
= {\rm Tr} \big( S_1^- \rho^{\rm ss} S_2^+ \big) & \neq & 0 ~.
\end{eqnarray}
Using Eq.~(\ref{reset}) and (\ref{215}) we obtain
\begin{eqnarray} \label{intensity2}
I_{\hat{\bf k}}(\rho^{\rm ss}) 
&=& {3 A \over 8 \pi} \left(1- | \hat{\bf D}_{21} \cdot \hat{\bf k} |^2 \right)
\nonumber \\
& & \times \Big[ {\rm Tr} \big( S_1^- \rho^{\rm ss} S_1^+ \big) 
+ {\rm Tr} \big( S_2^- \rho^{\rm ss} S_2^+ \big) \nonumber \\
& & + 2 \, {\rm Re} \, {\rm Tr} \big( 
{\rm e}^{-{\rm i}{\bf k}_0 \cdot ({\bf r}_1-{\bf r}_2)} 
S_1^- \rho^{\rm ss} S_2^+ \big) \Big]~,
\end{eqnarray}
where the last term describes the interference effects.

\noindent
\begin{minipage}{3.38truein}
\begin{center}
\begin{figure}[h]
\epsfig{file=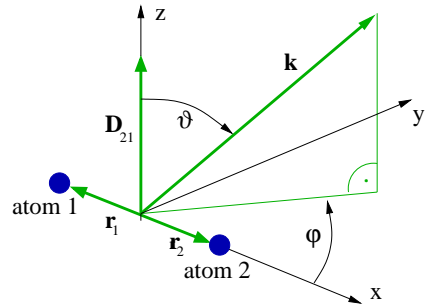,width=5.5cm}\\[0.3cm]
\caption{Coordinate system with the spatial angles $\vartheta$ and $\varphi$
characterising the direction of the wave vector ${\bf k}$. We assume 
that the atomic dipole moment 
${\bf D}_{21}$ is perpendicular to the line connecting both atoms.} 
\label{kos}
\end{figure}
\end{center}
\end{minipage}
\\[0.4cm]

To discuss a concrete example, it is convenient to introduce Dicke states,
\begin {eqnarray} \label{as_basis}
&& |g\rangle =  |11\rangle ~,~
|s\rangle =  \left(|12\rangle+|21\rangle\right)/\sqrt{2} ~, \nonumber\\ 
&& |e\rangle = |22\rangle ~,~
|a\rangle = \left(|12\rangle-|21\rangle\right)/\sqrt{2} ~,
\end {eqnarray}
and to use the spatial angles $\vartheta$ and $\varphi$ as defined in Fig.~\ref{kos}.
In the following we choose the dipole moments ${\bf D}_{21}$ to be
perpendicular to the line connecting both atoms.
Using this notation we find from Eq.~(\ref{intensity2}) in good agreement with 
Ref.~\cite{Rudolph}
\begin{eqnarray} \label{int-anti}
I_{\hat{\bf k}}(\rho^{\rm ss}) 
&=& \frac{3 A}{8 \pi} \, \sin^2\vartheta \,
\big[ 2\rho_{ee}+\rho_{ss}+\rho_{aa} \nonumber\\
&& + \left(\rho_{ss}-\rho_{aa}\right)
\cos\left(k_0 r \sin\vartheta\cos\varphi\right) \nonumber\\
&& + 2 \,{\rm Im} \, \rho_{sa} \sin\left(k_0 r \sin\vartheta \cos\varphi \right) \big]~,
\end{eqnarray}
where $\rho_{xy} \equiv \langle x|\rho^{\rm ss}| y \rangle$ are the matrix elements
of the steady state density matrix $\rho^{\rm ss}$. The last two terms in Eq. 
(\ref{int-anti}) describe the interference and result from the last term in 
Eq.~(\ref{intensity2}).

In the classical double-slit experiment, interference only occurs if the waves 
emanating from both slits have a stable phase relation. The same is true
for the phase difference of the Rabi frequencies driving both atoms. 
It enters Eq.~(\ref{int-anti}) through the steady state matrix $\rho^{\rm ss}$. 
As an example, we assume in the following that both atoms see the same (real) Rabi frequency
\begin{eqnarray} \label{Omega}
\Omega^{(1)} = \Omega^{(2)} = \Omega ~.
\end{eqnarray}
From Eq.~(\ref{master}) and the condition $\dot{\rho}^{\rm ss}=0$ we find 
\begin{eqnarray} \label{steady(2atom)}
&& \rho_{gg} = \frac{\left(A^2+\Omega^2\right)^2 
+ A^2 \left(2A + {\rm Re} \, C\right) {\rm Re} \, C
+ A^2 \left( {\rm Im} \, C\right)^2 }{N} ~, \nonumber\\
&& \rho_{ss} = \frac{\Omega^2 (2A^2 + \Omega^2)}{N} ~,~
\rho_{ee}  =  \rho_{aa} = \frac{\Omega^4}{N} ~,~ {\rm Im} \, \rho_{sa}  = 0 
\nonumber\\
\end{eqnarray}
with
\begin{equation} \label{N}
N = \left(A^2 + 2 \Omega^2 \right)^2 + A^2 (2A + {\rm Re} \, C) {\rm Re} \, C 
+ A^2 ({\rm Im} \,C)^2 ~. 
\end{equation}
As it can be seen from these equations,
the dipole interaction between the atoms has only a small influence 
on the depth but does not affect the form of the interference pattern.
For $r > 2 \, \lambda_0$ one can neglect all terms proportional to the dipole 
coupling constant $C$. This leads to \cite{note2}
\begin{eqnarray} \label{ifinal}
I_{\hat{\bf k}}(\rho^{\rm ss}) 
&=& \frac{3}{4\pi}  \, \frac{A \Omega^2}{\left( A^2 + 2 \Omega^2 \right)^2} 
\, \sin^2\vartheta \nonumber\\
& & \times \left[A^2+ 2 \Omega^2 + A^2 
\cos\left(k_0 r\sin\vartheta\cos\varphi\right)\right] 
\end{eqnarray}
which is in good agreement with experimental results by 
Eichmann {\em et al.}~\cite{Eichmann}.

\noindent
\begin{minipage}{3.38truein}
\begin{center}
\begin{figure}[h]
\epsfig{file=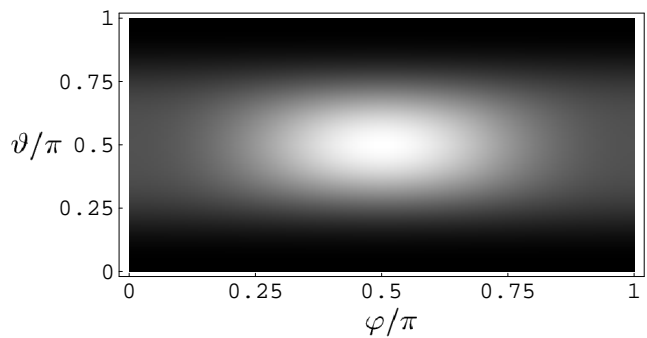,width=3.38truein}\\[0.3cm]
\caption{Density plot of the emission rate $I_{\hat{\bf k}}(\rho^{\rm ss})$
for two continuously driven two-level atoms, $r = \lambda_0 / \pi$ and 
$\Omega = 0.3 \, A$. White areas correspond to spatial angles with maximal 
intensity.} \label{int1}
\end{figure}
\end{center}
\end{minipage}

\noindent
\begin{minipage}{3.38truein}
\begin{center}
\begin{figure}[h]
\epsfig{file=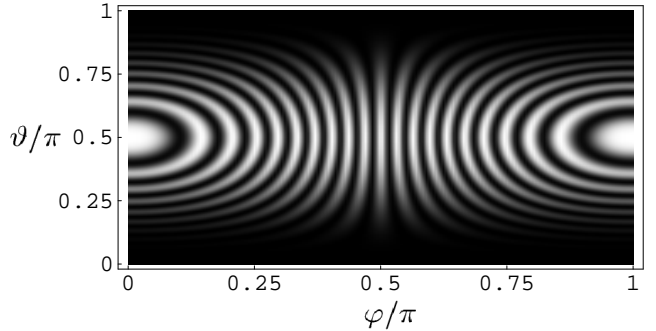,width=3.38truein}\\[0.3cm]
\caption{As in Fig.~\ref{int1} but with $r = 10 \, \lambda_0 $.} \label{int2}
\end{figure}
\end{center}
\end{minipage}
\\[0.4cm]

To illustrate this we show in Fig.~\ref{int1} and \ref{int2} density plots of 
the emission rate $I_{\hat{\bf k}}(\rho^{\rm ss})$ for different atomic distances $r$.
White areas correspond to spatial angles with maximal intensity.
The interference effects of the photons emitted by the 
two atoms are more distinct in Fig.~\ref{int2} which shows stronger oscillations 
of the intensity with the polar angle $\varphi$. These become more 
frequent the larger the distance between the atoms.

Finally we note that every change of the stationary state $\rho^{\rm ss}$ in
Eq.~(\ref{intensity2}) effects the spatial dependence of the interference pattern.
This has been discussed in Refs.~\cite{Kochan,Yeoman} where an additional
coupling of the two atoms via the mode of an optical cavity has been assumed. 
Another situation, in which the density matrix $\rho^{ss}$ is different from 
Eq.~(\ref{steady(2atom)}) is when the atomic state is continuously monitored. This can be 
done with the help of an additional rapidly decaying level and a second laser field
\cite{Scully82,Ingraham} or by using two four-level atoms and detecting the polarisation 
of the emitted photons \cite{Eichmann,Itano}. Alternatively, it has been proposed to 
use two microwave cavities as which way detectors \cite{Scully91}.
As a consequence of the knowledge of the which way information in these setups
the interference vanishes.
This in good agreement with our discussion in the previous subsection. 

\section{Bunching effects in the photon statistics of two distant atoms}

As another application of the quantum jump approach we investigate in this section 
the second order correlations in the photon statistics of two continuously driven 
two-level atoms. The experimental setup we consider is again the same as in 
Fig.~\ref{exp} but in the following we replace the screen by a single photon detector 
which registers only photons emitted in a certain direction $\hat{\bf k}$. 
In this section we predict strong spatially dependent {\em bunching} --- the 
effect that a photon emission in the $\hat{\bf k}$ direction increases the 
probability density for yet another emission in the same direction \cite{meystre}. 
Our results are in good agreement with the results of Ref.~\cite{Skornia}. An 
intuitive explanation for bunching, following the reasoning of Ref.~\cite{Beige},
is given.

To obtain a simple mathematical description of bunching we define, analogously to 
Eq.~(4) of Ref.~\cite{Beige}, the second order correlation function by
\begin{equation} \label{g0}
g^{(2)}_{\hat{\bf k}}(0) \equiv \frac{ I_{\hat{\bf k}}\left( 
R_{\hat{\bf k}}\,\rho^{\rm ss} \,R_{\hat{\bf k}}^{\dagger}/{\rm Tr}(\cdot) \right)}
{I_{\hat{\bf k}}(\rho^{\rm ss})} ~.
\end{equation}
The denominator of this function is the steady state photon emission rate
in the $\hat{\bf k}$ direction while the numerator equals the probability density 
for an emission in the same direction {\em immediately} after an emission. Therefore
the photons emitted in the $\hat{\bf k}$ direction are bunched if 
$g^{(2)}_{\hat{\bf k}}(0)>1$ 
and antibunched if $g^{(2)}_{\hat{\bf k}}(0) \le 1$.

\subsection{The photon correlation function for two continuously driven atoms}

With present ion trapping technology atomic distances larger than a few wave-lengths 
are easier to prepare. We consider therefore in the following 
the case $r > 2 \, \lambda_0$ and neglect again the dipole interaction between 
the atoms. Assuming, as in Eq.~(\ref{Omega}), that the Rabi frequency of the 
driving laser field is the same for both atoms we find from Eq.~(\ref{reset}),
(\ref{215}), (\ref{steady(2atom)}) and (\ref{N})
\begin{eqnarray} \label{g(0)}
g^{(2)}_{\hat{\bf k}}(0)  &=& 
\left[1-\frac{\cos\left(k_0r\sin\vartheta\cos\varphi\right)}
{1+2\left(\frac{\Omega}{A}\right)^2+\cos\left(k_0r\sin\vartheta\cos\varphi\right)}
\right]^2 ~.
\end {eqnarray}
As can be seen from this result, bunching occurs for all directions $\hat{\bf k}$ 
with $\cos\left(k_0r\sin\vartheta\cos\varphi\right) < 0$ and does not depend on the 
concrete choice of the Rabi frequency $\Omega$.
This is different from the statistics of photons emitted into {\em all} spatial 
directions where bunching can only occur for distances with 
$r < 2 \, \lambda_0$ \cite{Beige}. 

\noindent
\begin{minipage}{3.38truein}
\begin{center}
\begin{figure}[h]
\epsfig{file=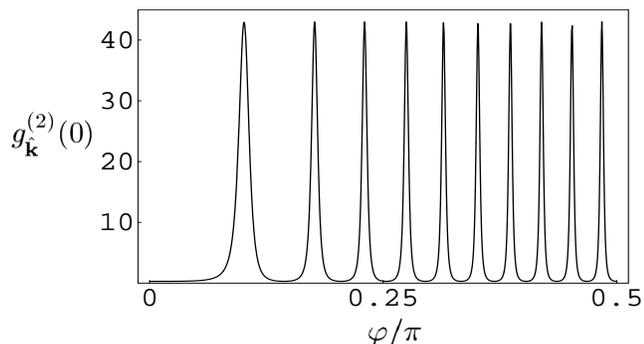,width=3.38truein}
\caption{The second order photon correlation function $g^{(2)}_{\hat{\bf k}}(0)$ 
as a function of $\varphi$ for $\Omega = 0.3 \, A$, $r = 10 \, \lambda_0$ and 
$\vartheta = \pi/2$.}
\label{cor1_p}
\end{figure}
\end{center}
\end{minipage}
\\[0.4cm]

Fig.~\ref{cor1_p} shows as an example the second order correlation function 
$g^{(2)}_{\hat{\bf k}}(0)$ for different spatial angles $\varphi$, $\vartheta = \pi/2$,
$r = 10 \, \lambda_0$ and $\Omega = 0.3 \, A$. For these parameters 
$g^{(2)}_{\hat{\bf k}}(0)$ 
can adopt values larger than $40$ which corresponds 
to very strong bunching. For weaker driving, $\Omega/A \to 0$, the correlation function 
can even become infinitely large. This seems unphysical but corresponds to 
angles for which the photon intensity (\ref{ifinal}) vanishes for $\Omega/A \to 0$.

\subsection{An intuitive explanation of strong bunching}

The quantum jump approach allows us not only to calculate easily
photon correlation functions but also to obtain a good intuitive understanding of 
this phenomenon. To do so we proceed as proposed in Ref.~\cite{Beige} 
and investigate how the state of the atoms changes during a photon emission 
in a direction with bunching. According to Eq.~(\ref{g(0)}) we get maximal bunching 
if
\begin{equation} \label{direction}
\cos\left( k_0 r \sin\vartheta \cos\varphi \right)=-1 ~.
\end{equation}
For this direction the corresponding reset operator (\ref{reset}) can be written as
\begin{eqnarray} \label{reset_bunching}
R_{\hat{\bf k}} & = & \alpha \left(|a\>\<e|-|g\>\<a|\right) ~,
\end{eqnarray}
where $\alpha$ is a complex number.
For the same direction the probability density for an emission (\ref {intensity}) 
equals
\begin{eqnarray} \label{Idir}
I_{\hat{\bf k}}(\rho^{\rm ss}) = |\alpha|^2 \left(\rho_{ee}+\rho_{aa}\right) 
\end{eqnarray}
and is proportional to the population in the states $|a \rangle$ and $|e \rangle$.

\noindent
\begin{minipage}{3.38truein}
\begin{center}
\begin{figure}[h]
\epsfig{file=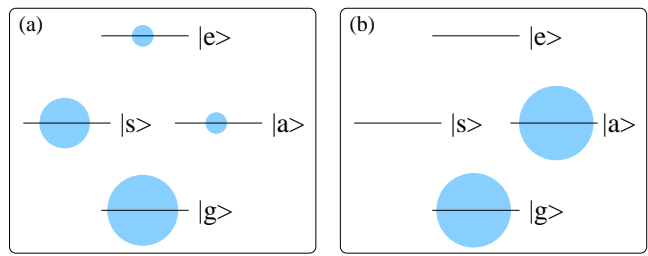,width=8.5cm}\\[0.3cm]
\caption{The population of the Dicke states $|g \>$, $|s \>$, $|a \>$ and $|e \>$
for the steady state $\rho^{\rm ss}$ (a) and for the normalised state immediately after the emission 
of a photon into a direction with maximal bunching (b) pictured by circles. The
area of each circle is proportional to the population of the corresponding level.}
\label{pop123}
\end{figure}
\end{center}
\end{minipage}
\\[0.4cm]

Fig.~\ref{pop123}(a) illustrates the population in the atomic levels
for the steady state matrix $\rho^{\rm ss}$ and Fig.~\ref{pop123}(b)
for the state of the atoms immediately after an emission in the direction 
$\hat{\bf k}$ of Eq.~(\ref{direction}). 
The area of each circle is proportional to the population of the 
corresponding level. In the steady state, there is nearly no population in
the levels $a$ and $e$ and the probability density for an emission in the 
direction of Eq.~(\ref{direction}) is therefore relatively low. It equals
\begin{equation} \label{anotherone}
I_{\hat{\bf k}}(\rho^{\rm ss})
= \frac{2\Omega^4|\alpha|^2}{\left(A^2+2\Omega^2\right)^2} ~.
\end{equation}
During an emission a redistribution of the population takes place according to 
the reset operator (\ref{reset_bunching}). The population of level $a$ goes over to 
level $g$ and the population of level $e$ goes to level $a$ while the population 
of the two other levels vanishes. Afterwards the reset state has to be normalised.
A comparison of Fig.~\ref{pop123}(a) and \ref{pop123}(b) shows that the emission 
of a photon causes in this way an increase of the population in the states $|a\>$ 
and $|e\>$ and therefore also an increase of the probability density for a further 
emission in the same direction, which is given by
\begin{eqnarray}
I_{\hat{\bf k}}\left( R_{\hat{\bf k}}\,\rho^{\rm ss}\,R_{\bf k}^{\dagger}
/{\rm Tr}(\cdot) \right)= {\textstyle{1 \over 2}} |\alpha|^2 ~,
\end{eqnarray}
which is larger than $I_{\hat{\bf k}}(\rho^{\rm ss})$ of Eq. (\ref{anotherone}).

Summarising this, we see that bunching results from the fact that the detection of 
a photon is always connected with a measurement on the atomic state. During this 
measurement the state of the atoms might change in such a way that the probability 
density for a further emission in the same direction is increased.

\section{Conclusions}

As long as a quantum mechanical system does not couple to its environment
one can predict its time evolution by the Schr\"odinger equation. 
This is not possible for open systems like spontaneously emitting atoms. 
To describe them the quantum jump approach \cite{Review} has been derived 
from the assumption that the environment performs continuous measurements 
on the free radiation field as to whether a photon is emitted by the atoms 
or not. The time evolution of the atoms under the 
condition of no photon emission can
be described by a Schr\"odinger equation based on the conditional Hamiltonian
$H_{\rm cond}$. In the case of a photon emission the state of the atoms 
changes abruptly. 

In this paper we assumed that the environment of the atom-field system, 
here in form of a screen, detects each emitted photon and, if so, 
determines its direction $\hat{\bf k}$. This ansatz was motivated by 
the experimental setup of Fig.~\ref{exp} in which each photon causes a 
``click'' at a certain point on the screen.
From this assumption of {\em environment induced measurements} we derived in 
Section II the reset operator $R_{\hat{\bf k}}$. It can be used to determine 
the state of the atoms immediately after an emission in the $\hat{\bf k}$ direction.
Initially in a pure state, the state of the atoms remains always pure.
This extension of the quantum jump approach allows us now to predict all individual 
trajectories of a single atomic system. We think that all quantum optical 
experiments with ``white'' walls in the laboratory can be described by a quantum 
jump approach. 

In Section III we showed that our results are consistent with the master 
equations for two dipole interacting atoms \cite{Agarwal}. 
The dipole interaction results from the fact that both atoms interact with the same 
free radiation field and exchange virtual photons. This
is described by the dipole coupling constant $C$ in the conditional 
Hamiltonian $H_{\rm cond}$. Also the reset operator $R_{\hat{\bf k}}$ 
leads to terms proportional $C$ in the master equation.

The advantage of our generalisation of the quantum jump approach \cite{Beige,Beige2} 
is that it can now be applied to further experiments such as the scheme by 
Cabrillo {\em et al.} \cite{Cabrillo} to entangle distant atoms by interference.
In this paper we discussed in Section IV, as an example, the two-atom double-slit 
experiment shown in Fig.~\ref{exp} and derived a necessary and sufficient 
interference criterion. 
Another application of the reset operator $R_{\hat{\bf k}}$
was given in Section V, where we predicted in agreement with Ref.~\cite{Skornia} strong 
bunching for the photons emitted into certain directions $\hat{\bf k}$. 
An intuitive explanation for this effect was given.

\narrowtext
{\em Acknowledgment.}
We would like to thank P. L. Knight, J.-L. Lehners, 
A. Loettgers and P. Millen for interesting and stimulating discussions.
This work is based on an essay written by C. S. during a student exchange 
program at Imperial College in London. He would like to express his gratitude 
for the hospitality he experienced at the Blackett Laboratory in the group of 
P. L. Knight. This work was also supported by the A. v. Humboldt Foundation and
by the European Union.

\end{multicols}

\end{document}